# A multi-milliJoule femtosecond Raman laser emitting at 1.28 μm.

CARLO VICARIO,[1,*] MOSTAFA SHALABY,[1] ALEKSANDR KONYASHCHENKO,[2] LEONID LOSEV[2], AND CHRISTOPH P. HAURI[1,3,*]

[1]SwissFEL, Paul Scherrer Institute, 5232 Villigen-PSI, Switzerland
[2] P.N.Lebedev Physical Institute, Leninsky prospect, 53, Moscow, 119991, Russia
[3]Ecole Polytechnique Federale de Lausanne, 1015 Lausanne, Switzerland

*carlo.vicario@psi.ch; christoph.hauri@psi.ch

**We report on the generation of broadband, high-energy femtosecond pulses centered at 1.28 μm by stimulated Raman scattering in pressurized hydrogen cell. Stimulated Raman scattering is performed by two chirped and delayed pulses originating from a multi-mJ Ti:Sapphire amplifier. The Stokes pulse carries energy of 4.4 mJ and is recompressed down to 66 fs by reflective grating pair. We characterized the short-wavelength mid-infrared source in view of energy stability, beam profile and conversion efficiency at a repetition rate of 100 Hz and 10 Hz. The demonstrated laser will benefit intense THz generation applications from highly nonlinear organic crystals.**

Nonlinear frequency down-conversion of intense femtosecond (fs) pulses from Ti:Sapphire lasers is an established concept in high-field research to provide pulses at wavelength longer than 1 μm. Mid-infrared fs laser sources have found many applications including the generation of attosecond pulses by high-order harmonic generation (HHG) [1] and the production of strong terahertz fields [2-11]. In particular, present intense THz source technology (with peak fields > 1 GV/m) depends on optical rectification of femtosecond pulses in highly nonlinear organic crystals. However, the main challenge in such systems is the requirements of high-energy non-standard pump laser wavelengths (1.2 – 1.5 μm). At those wavelengths the optical rectification process presents best phase matching conditions. Presently Ti:sapphire-driven optical parametric amplifiers at typical 100 Hz repetition rate are the most common approach for shifting the spectral content to longer wavelengths [12]. This multi-stage nonlinear conversion, however, suffers typically from limited pulse energy, reduced conversion bandwidth and an inhomogeneous beam profile. The latter is particularly detrimental for driving THz generation with organic crystals as hot spots can induce fatal damages.

Another approach for high-energy THz generation in highly nonlinear organic crystal relies on Cr:fosterite laser technology [13]. With such a laser pump, THz pulse energy up to 900 μJ and electric field of 50 MV/cm have been demonstrated in different organic nonlinear media [7, 8]. This technique offers significantly better THz beam profile than the OPA one. This gives it a significant advantage in handling and focusing the beams in complex pump-probe spectroscopy [3, 6, 7]. However, so far the THz source driven by Cr:Fosterite laser has been demonstrated only at 10 Hz [4-5] and therefore it suffers from much lower signal to noise ratio than the OPA approach.

In this letter we present efficient frequency down-conversion based on stimulated Raman scattering (SRS) in gas that combines the homogenous beam profile of the Cr:fosterite approach and the high repetition rate of the OPA one. Moreover for the Raman laser the phase matching bandwidth supports ultrashort pulses and the converted pulse energies can potentially reach several mJ level. Our Raman laser operates with uncompressed pulses from a Ti:Sapphire laser with energy of 30 mJ and repetition rate of 100 Hz. To our knowledge, the present source is the most advanced fs Raman laser scheme in terms of average power, peak intensity and pulse energy delivering femtosecond pulses at 100 Hz repetition rate.

SRS sources for the generation of intense femtosecond pulses, especially at wavelengths longer 1 μm, have not been extensively studied in the past. Grigsby et al. reported maximum pulse energy of the Stokes radiation at 870 nm of 3 mJ using a 10 Hz Ti:sapphire laser pump for SRS in a barium nitrate crystal. This resulted in pulse duration of 115 fs [14]. On the other hand SRS has also been explored for the production of ultrashort pulses. The shortest Stokes pulse with a duration of 40 fs was generated in high pressure hydrogen cell pumped by a 35 fs short Ti:Sapphire laser at 20 Hz. The first Stokes radiation at a wavelength of 1.25 μm carries moderate pulse energy of 0.4 mJ [15]. A 50-fs hydrogen Raman laser with pulse repetition rate of 2 kHz was recently demonstrated by using quartz capillary as an optical waveguide. The Stokes pulse energy was in this case ∼ 10 μJ [16]. In all the cited works the energy conversion efficiency from the fundamental pump wavelength to the first Stokes component was varying between 10-20%.

If femtosecond laser pulses are applied, the main difficulty for upscaling the SRS approach to higher energies is the onset of unwanted nonlinear effects, such as self-phase modulation (SPM) and self-focusing (SF). An alternative configuration to overcome those limitations is to operate the femtosecond Raman laser scheme by a chirped double-pulse pumping to avoid the narrowing of the Stokes pulse spectrum [9]. In order to reduce the SPM and SF competing with the SRS process, the advanced SRS scheme used here employs chirped laser pulses with duration of about two hundred picoseconds, similar to [8]. Such pulses are provided directly from a chirped pulse amplifier system without employing the final grating compressor. The chirped Stokes pulse undergoes a spectral shift to longer wavelength by SRS in a gas cell and is then compressed by a grating compressor to the femtosecond regime.

The hydrogen Raman laser setup and the main experimental parameters are reported in Fig.1. The driving laser consists of a Ti:Sapphire oscillator and a multistage chirped pulse amplifier delivering 30 mJ pulses at 10 and 100 Hz [17]. As said before, in order to avoid detrimental effects induced by high peak power, 200 ps long pulses are used to pump the Raman cell.

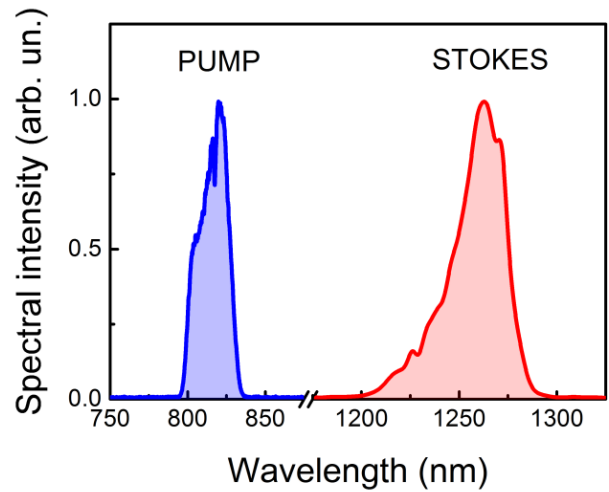

**Fig. 2.** The high-energy $H_2$ Raman laser allows to redshift the Ti:Sa laser spectrum centered at 817 nm (blue curve) to a spectrum centered at 1270 nm (red curve). The spectral shape and the width is mainly preserved in the SRS process.

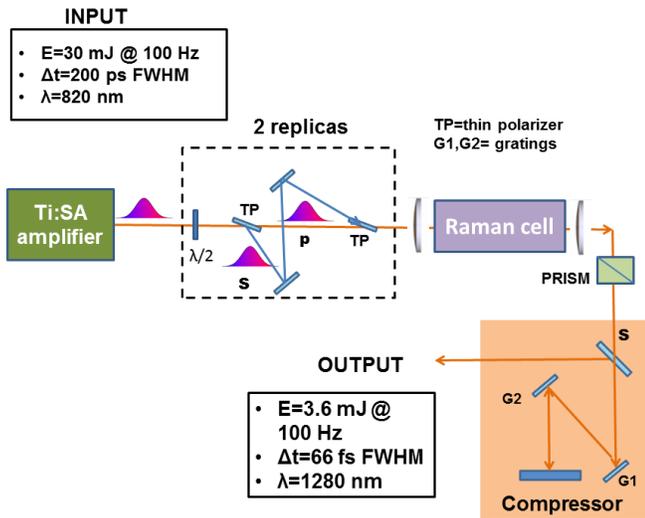

**Fig. 1** Optical layout of the high-energy $H_2$ Raman laser. The laser pump and output parameter are also reported.

The Ti:Sapphire laser amplifier includes DAZZLER and MAZZLER (by Fastlite) acousto-optical programmable filters for advanced spectral amplitude and phase shaping control [17]. This layout allows us to tune the amplifier peak wavelength and the spectral width and optimize the final SRS pulse duration by spectral phase shaping. For the present measurements the laser central wavelength is set at 817 nm with a FWHM bandwidth of 30 nm.

Two orthogonally polarized pulses delayed by about 250 ps are produced prior to the Raman cell using two thin film polarizers. The ratio between the energy of the two pulses can be varied. We found that the maximum conversion efficiency is obtained at the energy of 17 mJ for the first and 13 mJ for the second pulse, respectively. When focused in the pressurized $H_2$ gas medium, the first pulse generates a coherent phonon wave on which SRS of the delayed replica occurs with high efficiency. The Raman cell is a 1.5 meter long chamber and it is filled with $H_2$ (up to 9 bar) and krypton. For all measurements krypton was mixed to $H_2$ to suppress detrimental nonlinear and parametric effects such as four wave mixing process which gives rise to anti-Stokes wavelength and decrease the Raman amplification. We found experimentally that krypton pressure of 2 bar allows for the maximum Raman scattering conversion efficiency. The Stokes radiation of the delayed pulse is separated after the Raman cell using a Rachon prism and is recompressed by means of a reflective grating compressor with efficiency of 80%. The maximum femtosecond Stokes pulse energy obtained before (after) compression in the experiment is 4.4 mJ (3.6 mJ). This is the largest pulse energy and highest average power ever reached in femtosecond hydrogen Raman laser. Considering the input energy of the amplified delayed pulse (13 mJ) this corresponds to a photon conversion of 53% (42% after the compressor). Whereas the total energy conversion efficiency is 12% after the compressor.

Shown in Figure 2 are the pump (blue curve) and Stokes pulse (red curve) spectra. The Stokes spectrum is centered at 1280 nm and has a bandwidth of 40 nm FWHM and 80 nm at $1/e^2$ level. The spectral shape of the Stokes pulse mimics approximatively the asymmetric shape of the Ti:Sa amplifier pump. The spectral width of the Stokes pulse shows only a small bandwidth reduction compared to the original pump laser (≈20% in the frequency scale) which indicates that our setup is capable to transfer most of the spectral contents from the pump to the first Stokes pulse.

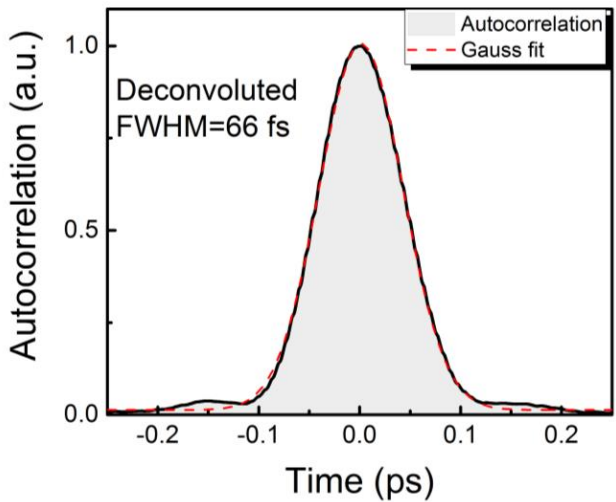

**Fig. 3.** The Stokes pulse centered at a wavelength of λ=1.25 μm is recompressed to 66 fs FWHM using a grating compressor. The pulse duration is optimized by adapting the input phase with the acousto-optical filter installed in the Ti:Sapphire amplifier. The pulse duration is measured with a single shot autocorrelator and is close to transform limit.

Downstream the Raman cell the Stokes pulse is compressed by mean of reflective grating pair. The overall energy transmission in this system is about 80%. The compressor mainly compensates for the linear chirp that is transferred from the pump laser to the SRS pulse. In addition the residual high order phase dispersion is compensated by the DAZZLER at the laser pump wavelength. The duration of the compressed Stokes pulse is measured by means of a single-shot autocorrelator (Minioptics). Reported in Fig. 3 is the autocorrelation trace of the compressed Stokes pulse. Assuming a Gaussian shape, the de-convoluted full width half maximum is 66 fs. This value is close to the transform limited duration for the given Stokes spectrum (65 fs) shown in Fig. 2. The measured Stokes pulse duration is 20% longer than the original pump laser pulse. The results are in line with the expected spectral narrowing occurring in the Raman nonlinear process. The pedestal in the autocorrelation trace indicates however residual high-order spectral phase.

Shown in Fig. 4 are the measured pulse energy and rms energy stability recorded after the compressor for 100 Hz (graphs a) and b)) and 10 Hz (plots c) and d)) pump repetition rate, respectively. The efficiency is calculated with respect to the total input laser energy distributed over the two pulses. These parameters are plotted as function of the hydrogen pressure. For all these measurements the Kr backing pressure was set at 2 bar. At 100 Hz (4a) the Raman pulse energy linearly scales with the $H_2$ pressure and reaches up to 3.6 mJ at the highest applicable pressure (9 bar). At this pressure the overall conversion efficiency reaches 12% (15% before compression).

The decreasing first Stokes pulse energy at high hydrogen pressure for 10 Hz pump repetition rate can be explained by energy conversion to second Stokes radiation with wavelength of about 2.5 μm [20]. Because of the Raman amplification is proportional to the concentration of the $H_2$ at transient SRS the conversion to the second Stokes grows with increasing hydrogen pressure. At 100 Hz pulse repetition rate the second Stokes appears at higher hydrogen pressure because of gas heating by the laser, in the interaction volume results in decreasing concentration of $H_2$ and the Raman amplification drops as consequence. The minimum in the rms Stokes energy stability (Fig. 4b and 4d) can be explained as following. At low hydrogen pressure the first Stokes pulse energy fluctuations grow because of the Raman amplification is close to the amplification threshold [21]. At higher hydrogen pressure the first Stokes rms energy stability increases due to the onset of the second Stokes generation. Shot to shot energy stability of 0.04 rms is achieved at 8 bar for 100 Hz and 6 bar for 10 Hz, respectively.

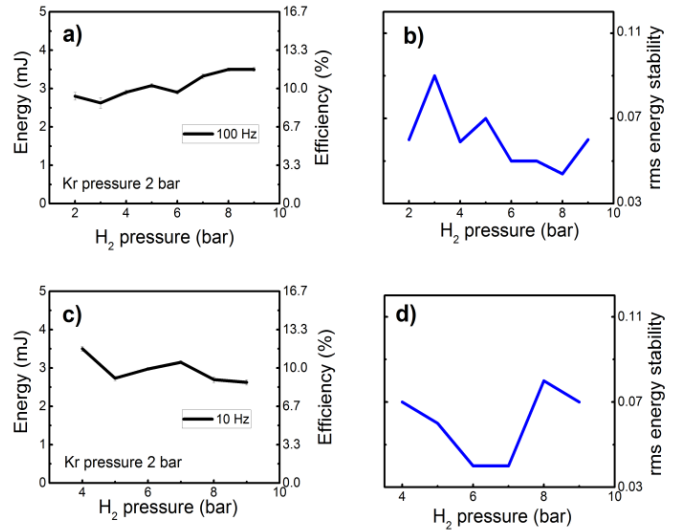

**Fig. 4.** a), c) The Stokes pulse energy and b), d) stability as function of the hydrogen pressure. Two laser repetition rates (10, 100 Hz) have been investigated.

As mentioned earlier, one of the main issues for high-energy optical parametric amplifiers is the irregular beam profile which often suffers from hot spots and significant optical aberrations. These features are particularly harmful for efficient THz generation by optical rectification in organic crystal where intensities close to the crystal damage threshold are applied in order to maximize the conversion efficiency. For the most common organic nonlinear materials, the relative low laser damage threshold (<20 mJ/cm$^2$) demands for uniform beam profile and absence of hot-spot. Moreover strong aberrations are detrimental for efficient high-harmonic generation in gas where good focus quality is needed. In the experimental setup the Stokes beam profile is measured after compression with a pyroelectric camera having pixel size of 100 μm (Ophir Photonics Pyrocam III). As shown in Fig. 5 for repetition rate of 100 and 10 Hz, the Stokes pulse presents a regular transverse intensity free of hot spot. The intensity profiles along the vertical and horizontal axes can be well approximated with Gaussian fit having for 100 Hz repetition rate b) FWHM of 2.7 and 2.8 mm and for 10 Hz FWHM of 1.9-1.7 mm a) respectively. The two beam profiles are recorded for $H_2$ pressure of 8 bar. For 10 Hz repetition rate the beam is more focused due probably to less gas heating. This explains also why the Raman gain differs for the two repetition rates. At 10 Hz the Stokes beam is smaller and does not fully take profit of volume of the gain medium.

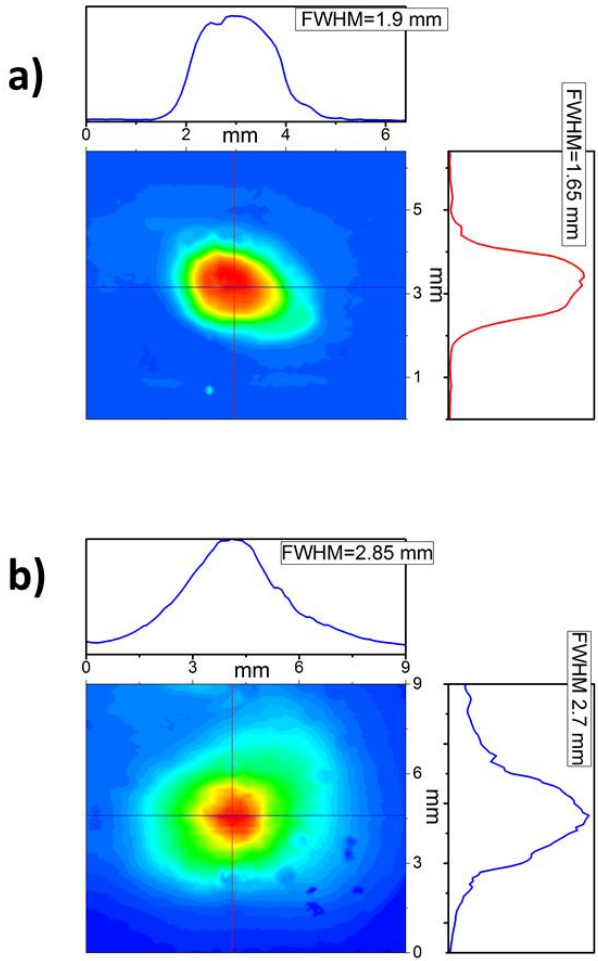

**Fig. 5** The beam profile of the compressed SRS pulse at a) repetition rate of 10 Hz and b) 100 Hz, respectively. The beam profile is free of intensity hot spots and exhibits a Gaussian-like shape.

In conclusion we presented high-power Raman laser based on chirped-pulse amplification in hydrogen delivering record-high pulse energy of 4.4 mJ at wavelength of 1.28 μm and repetition rate of 100 Hz. This corresponds to a photon conversion efficiency exceeding 50 %. The present results represent the largest ever-reported pulse energy and power for hydrogen stimulated Raman scattering laser. The chirped Stokes pulse is recompressed to 66 fs by means of a grating compressor with maximum pulse energy of 3.6 mJ. The resulting intense mid-infrared femtosecond pulses will find applications in extreme nonlinear optics and strong-field THz sources based on organic crystals [8], as the required pump wavelength coincides with the wavelength of the first Stokes component of SRS of Ti:Sapphire laser in hydrogen.

**Acknowledgment:** We are grateful to Marta Divall, Alexandre Trisorio, and Andreas Dax from the SwissFEL laser group for supporting the operation of the Ti:Sapphire laser system. We acknowledge the support from Edwin Divall in DAQ. We acknowledge financial support from the Swiss National Science Foundation (SNSF) (Grant No. 200021_146769 and No. IZLRZ2-164051). MS acknowledges partial funding from the European Community's Seventh Framework Programme (FP7/2007-2013) under grant agreement no. 290605 (PSI-FELLOW/COFUND). CPH acknowledges association to NCCR-MUST.